\documentclass[preprint,prl,superscriptaddress,amsmath,amssymb]{revtex4}

\usepackage{bm}
\usepackage{color}
\usepackage[dvips]{graphicx}

\usepackage{bm}

\begin{document}

\title{Strongly spin-orbit coupled two-dimensional electron gas emerging near the surface of polar semiconductors}

\author{M. \surname{Sakano} }
\affiliation{Department of Applied Physics, The University of Tokyo, Tokyo 113-8656, Japan}

\author{M. S. \surname{Bahramy} }
\affiliation{Correlated Electron Research Group (CERG), RIKEN Advanced Science Institute, Wako, Saitama 351-0198, Japan}

\author{A. \surname{Katayama} }
\affiliation{Department of Applied Physics, The University of Tokyo, Tokyo 113-8656, Japan}

\author{T. \surname{Shimojima} }
\affiliation{Department of Applied Physics, The University of Tokyo, Tokyo 113-8656, Japan}

\author{H. \surname{Murakawa} }
\affiliation{Correlated Electron Research Group (CERG), RIKEN Advanced Science Institute, Wako, Saitama 351-0198, Japan}

\author{Y. \surname{Kaneko} }
\affiliation{Correlated Electron Research Group (CERG), RIKEN Advanced Science Institute, Wako, Saitama 351-0198, Japan}

\author{W. \surname{Malaeb} }
\affiliation{Institute of Solid State Physics, The University of Tokyo, Kashiwa, Chiba, 227-8581, Japan}
\affiliation{CREST,JST, Tokyo 102-0075, Japan}

\author{S. \surname{Shin} }
\affiliation{Institute of Solid State Physics, The University of Tokyo, Kashiwa, Chiba, 227-8581, Japan}
\affiliation{CREST,JST, Tokyo 102-0075, Japan}

\author{K. \surname{Ono} }
\affiliation{Institute of Materials Structure Science, High Energy Accelerator Research Organization (KEK), Tsukuba, Ibaraki 305-0801, Japan}

\author{H. \surname{Kumigashira} }
\affiliation{Institute of Materials Structure Science, High Energy Accelerator Research Organization (KEK), Tsukuba, Ibaraki 305-0801, Japan}

\author{R. \surname{Arita} }
\affiliation{Department of Applied Physics, The University of Tokyo, Tokyo 113-8656, Japan}
\affiliation{Correlated Electron Research Group (CERG), RIKEN Advanced Science Institute, Wako, Saitama 351-0198, Japan}
\affiliation{JST, PRESTO, Kawaguchi, Saitama 332-0012, Japan}

\author{N. \surname{Nagaosa} }
\affiliation{Department of Applied Physics, The University of Tokyo, Tokyo 113-8656, Japan}
\affiliation{Correlated Electron Research Group (CERG), RIKEN Advanced Science Institute, Wako, Saitama 351-0198, Japan}

\author{H. Y. \surname{Hwang}}
\affiliation{Correlated Electron Research Group (CERG), RIKEN Advanced Science Institute, Wako, Saitama 351-0198, Japan}
\affiliation{Department of Applied Physics and Stanford Institute for Materials and Energy Science, Stanford University, 
Stanford, California 94305, USA}

\author{Y. \surname{Tokura} }
\affiliation{Department of Applied Physics, The University of Tokyo, Tokyo 113-8656, Japan}
\affiliation{Correlated Electron Research Group (CERG), RIKEN Advanced Science Institute, Wako, Saitama 351-0198, Japan}

\author{K. \surname{Ishizaka} }
\affiliation{Department of Applied Physics, The University of Tokyo, Tokyo 113-8656, Japan}
\affiliation{JST, PRESTO, Kawaguchi, Saitama 332-0012, Japan}

\begin{abstract}
We investigate the two-dimensional (2D) highly spin-polarized electron accumulation layers commonly appearing near the surface of 
$n$-type polar semiconductors BiTeX (X = I, Br, and Cl) by angular-resolved photoemission spectroscopy.
 Due to the polarity and the strong spin-orbit interaction built in the bulk atomic configurations, the quantized conduction-band subbands  
show giant Rashba-type spin-splitting. 
The characteristic 2D confinement effect is clearly observed also in the valence-bands down to the binding energy of 4 eV. 
The X-dependent Rashba spin-orbit coupling is directly estimated from the observed spin-split subbands, 
which roughly scales with the inverse of the band-gap size in BiTeX. 
\end{abstract}

\maketitle

In a nonmagnetic solid compound, spatial-inversion and time-reversal symmetries require the degeneracy of an electron's spin, which 
can be lifted in a noncentrosymmetric system by introducing spin-orbit interaction (SOI). 
The Bychkov-Rashba (BR) effect describes such SOI in an inversion-symmetry broken system, by simply considering a two-dimensional (2D) electron gas with an electric field applied along its perpendicular direction \cite{Bychkov84}. 
The Hamiltonian of the BR model is given by $H_{\rm{BR}}=\lambda \, \vec{\sigma} \cdot (\vec E_z \times \vec{k})$, where $\lambda$ is the coupling constant reflecting the SOI, $\vec{\sigma}$ and $\vec k$ are the spin and momentum of electrons, and $\vec E_z$ is the applied electric field.
It gives an the effective magnetic field acting on the electron spin $\vec{\sigma}$, whose orientation is normal to $\vec E_z$ and $\vec k$. 
The energy dispersion is characterized by the $k$-linear splitting of the free electron parabolic band dispersion as denoted by $E^\pm (\vec{k}) = \hbar ^2 {\vec{k}}^2/2m^{\ast} \pm {\alpha}_{\rm{R}} |\vec{k}|$, where $m^{\ast}$ is the effective mass of the electron and $\alpha_{\rm{R}}$ gives the degree of Rashba-type spin-splitting $i.e.$ Rashba parameter. 
Until now, such asymmetry-induced spin-splitting has been actively investigated and designed in semiconductor heterostructures such 
as the InGaAs/GaAs system \cite{Nitta97}. 
More recently, attempts to induce spin-splitting by applying the electric field in centrosymmetric heavy-element oxide semiconductors, such as KTaO$_3$, are also reported \cite{Nakamura09}. 
The spin-splittings in these systems have remained small compared to the thermal energy at room temperature, nevertheless, due to the weak SOI and/or small asymmetric $\vec E_z$ that can be applied by such field effects. 

Previously we reported the finding of a giant Rashba-type spin-splitting realized in the bulk of a polar semiconductor, BiTeI \cite{Ishizaka11,Bahramy11,Sakano12,Lee11}. 
Its peculiar polar crystal structure leads to a bulk-induced Rashba-type SOI, similarly as originally discussed in wurzite semiconductor materials \cite{Rashba60}. 
The Rashba parameter ${\alpha}_{\rm{R}}$ for BiTeI is found to be as large as $3.5$ eV\AA , which is comparable to the largest value reported thus far for a surface alloyed metal Bi/Ag(111) system \cite{Ast07}. 
By utilizing such a large Rashba-type spin-splitting nature built in the bulk semiconductor material, one may obtain a 2D electron system under a strong Rashba field near its surface and interface, which has been highly desired from the viewpoint of developing various spintronic functions \cite{Datta90,Sinova04} and multi-functional interfaces \cite{Hwang12}. 
In this Letter, we investigate the near-surface regions of the extended families of polar semiconductors BiTeX (X=I, Br, and Cl) by means of high-resolution angular-resolved photoemission spectroscopy (ARPES). 
Using first-principles calculations combined with the Poisson-Schr\"{o}dinger (PS) method \cite{King08}, we precisely characterize the 2D 
confinement effect on electronic structures including both conduction- and valence-bands, emerging as a consequence of charge accumulation at the surfaces.


Single crystals of BiTeI were grown by the Bridgman method \cite{Ishizaka11}, whereas those of 
BiTeBr and BiTeCl were grown by the chemical vapor transport method. 
Laser-ARPES was performed using a vacuum ultraviolet laser ($h\nu = 6.994$ eV) and a VG-Scienta R4000 electron analyzer \cite{Kiss08} 
with a total resolution energy of 4 meV. 
$h\nu$-dependent ARPES measurement ($h\nu = 60$ - $84$ eV) was performed by using the VG-Scienta SES2002 electron analyzer at BL28 in Photon Factory, KEK,  
with a total resolution energy of $20$ - $30$ meV. 
Samples were cleaved {\it in situ} at around room temperature and measured at 10-15 K. 
Electronic-structure calculations for BiTeX were carried out within the context of density functional theory using the full-potential augmented plane-wave plus local orbital (APW-LO) method \cite{Ishizaka11}. 
To simulate the effect of band bending, a 60 layer tight-binding supercell was constructed by downfolding the APW-LO Hamiltonian, using maximally localized Wannier functions \cite{souza,mostofi,kunes}. 
The band bending potential was then obtained by solving the coupled PS equation [13], assuming static dielectric constants 15, 10, and 6.5 for X=I, Br, and Cl, respectively.

The crystal structures of BiTeX are shown in Figs. 1(a) (X = I and Br; space group $P3m1$) and 1(b) (X = Cl; $P6_{3}mc$).
They are characterized by the alternative stacking of Bi-, Te- and X-layers, which gives rise to the polarity along the stacking axis. 
In contrast to the previous report indicating the disorder of Te and Br sites for BiTeBr \cite{Shevelkov95}, 
the single crystalline BiTeBr we obtained was confirmed to possess the well-defined stacking of Te- and Br- triangular lattices by x-ray diffraction measurements.

The calculated band structures for BiTeI and BiTeBr (BiTeCl) as shown in Fig. 1(f)-1(h) are composed of $6$ ($12$) conduction bands above the Fermi level ($E_{\rm{F}}$) and $12$ ($24$) valence bands below $E_{\rm{F}}$. 
The conduction bands for BiTeX are predominantly composed of Bi $6p$ components. 
The valence bands, on the other hand, show a systematic X-dependence, as seen by the lower half of them moving toward deeper binding energy ($E_{\rm B}$) upon increasing the electronegativity of X from I, Br, to Cl. 
This indicates that the lower set of the valence bands is mainly dominated by X ions, whereas the upper ones primarily reflect the Te $5p$ character. 
The resistivity for each sample, as shown in Fig. 1(e), indicates metallic behavior down to $2$ K. 
The electron densities of X=I, Br, and Cl samples are estimated from the Hall coefficient to be $n_{\rm{bulk}}= 4 \times 10^{19}$, $4 \times 10^{18}$, and $2 \times 10^{18}$ cm$^{-3}$, respectively, indicating that they all behave as $n$-type degenerate semiconductors due to slight nonstoichiometry. 
From the band calculation, the Fermi levels of the bulk state corresponding to these carrier concentrations should be located at
 $142$ meV (X=I), $27$ meV (X=Br), and $5$ meV (X=Cl) above their respective conduction band minimums (CBMs).

Figure 2(a) shows the laser-ARPES image for BiTeBr, recorded along the $\Gamma$-K ($k_y$) direction. 
The CBM reaches $E_{\rm B} \sim 0.47$ eV, with the remarkable Rashba-type band splitting of up to $\sim 0.2$ eV. 
The observed CBM is quite deep below $E_{\rm F}$, considering that the expected Fermi level in the bulk BiTeBr should be located at merely $\sim 27$ meV above the CBM at $k_z=\pi /c$. 
One can also notice that there are two additional sets of bands at shallower energies with CBMs at around 
$E_{\rm B} = 0.2$ and $0.1$ eV, 
thus forming a ladder of three Rashba-split subbands. 
These clearly indicate that downward band bending occurs near the surface of BiTeBr, forming an $n$-type accumulation layer. 
The conduction electrons confined in such 2D accumulation layers are known to form quantized conduction subbands \cite{King08,Santander11,Meevasana11}. 
In the present case of BiTeBr, all the subbands clearly distinguished indicate the giant Rashba-like splittings, thus offering direct evidence of a 2D spin-polarized electron system dominated by the huge Rashba-type effective magnetic field.

To grasp the topology and shape of these spin-split subbands,  
the constant energy cross-sectional views of the band dispersions are shown in Fig. 2(b). 
At $E_{\rm B} =0$, $i.e.$ $E_{\rm{F}}$, six Fermi surfaces (FSs) corresponding to inner and outer FSs for the $n= 1 \sim 3$ subbands are observed. 
The large ones, namely the outer and inner $n=1$ FSs, are strongly distorted compared to the other small circular FSs.
Such anisotropy arises from the higher-order terms of SOI reflecting the underlying crystal symmetry, beyond the Rashba-type ($k$-linear) spin-splitting, that becomes appreciable at large $k$. 
At deeper $E_{\rm B}$, the cross-sections of the bands become smaller and more isotropic. 
The band crossing point (Rashba point) is also clearly distinguished as a sharp point appearing at the zone center for some specific $E_{\rm B}$ ({\it e.g.} $160$ meV for the $n=2$ subband). 
To discuss the realized 2D electronic structure, we introduce the PS method to calculate the subband structures by taking into account the band bending potential and 2D confinement at the accumulation layer \cite{King08}. 
Here, we fix the bulk carrier density and calculate the depth ($z$)-dependent energy $V(z)$ and electron density $n(z)$, which most accurately reproduce the observed subbands configuration [see Fig. 3(b)]. 
$V(z)$ and $n(z)$ are shown in Figs. 2(c) and (d), which indicate that the electron accumulation layer of $\sim 1$ nm thickness is formed at the near-surface region of BiTeBr.

The systematic results of laser-ARPES and corresponding PS simulation for BiTeX are shown in Fig. 3.  
The experimentally obtained subband dispersions along $k_y$ directions are shown in Figs. 3(a)-3(c). 
Markers showing the peak positions of ARPES intensity estimated from the energy distribution curves and momentum distribution curves are overlaid on the images. 
BiTeI (BiTeBr and BiTeCl) exhibits two (three) sets of Rashba-type spin-split conduction subbands. Subbands of $n = 1$, $2$, and $3$ are depicted by red, blue, and green markers. 
The CBMs of the deepest subband ($n=1$) are located at $E_{\rm B} \sim 0.27$, $0.47$, and $0.41$ eV for X=I, Br, and Cl, respectively, which are all much deeper when compared to the respective bulk $E_{\rm{F}} - E_{\rm CBM}$ ($0.142$, $0.027$, and $0.005$ eV).
The 2D electron density at the accumulation layer can be estimated experimentally by counting the areas of all the observed FSs in Figs. 3(g)-3(i); $n_{\rm{2D}} = 3.0 \times 10^{13}$, $7.1 \times 10^{13}$, and $4.6 \times 10^{13}$ cm$^{-2}$ for X=I, Br, and Cl, respectively.
The $V(z)$ and $n(z)$ which most well reproduce the observed subband positions for BiTeX are given in Figs. 2(c) and 2(d). 
The calculated subband dispersions in Figs. 3(d)-3(f) agree well with the ARPES data, including the group velocities and spin-splittings of the respective subbands. 
By comparing the FSs [Figs. 3(g)-3(l)], we can also demonstrate that the PS method quantifies their sizes and shapes ($k_x$-$k_y$ anisotropy) as well. 
Note that the $n=1$ inner and outer FSs get closer at $k_x$ ($\Gamma$-M) direction as compared to $k_y$ ($\Gamma$-K) direction. 
This behavior can be qualitatively understood by looking at the bulk band dispersions near the CBM. 
For BiTeBr as shown in Fig. 1(g), for example, the spin-split band dispersions along the A-H ($k_y$) direction remain almost parallel, whereas those along A-L ($k_x$) cross each other at $\sim 0.8$ eV above the CBM due to the trigonal warping effect, that dominates at large $k$ vectors \cite{Bahramy12}. 
This anisotropy gives rise to the observed deformation of the FSs.

To quantitatively evaluate the spin-splittings, we estimate the Rashba-parameter $\alpha_{\rm{R}}$ for BiTeX. 
Here we calculate $\alpha_{\rm{R}}$ by using the momentum offset at CBMs ($k_0$) and the Rashba energy ($E_{\rm{R}}$), as depicted in Fig. 3(d). 
They are related by $E_{\rm{R}} = \hbar ^2 {k_0}^2/2m^{\ast}$ and ${\alpha}_{\rm{R}} = 2E_{\rm{R}}/k_0$ in the 2D free-electron BR model. 
Estimated values for $E_{\rm{R}}$, $k_0$ and $\alpha_{\rm{R}}$ are shown in Table I. 
There is no clear $n$-dependence in the obtained $\alpha_{\rm{R}}$ ($e.g.$ for BiTeBr, $\alpha_{\rm{R}} = 2.0 \pm 0.7$, $2.0 \pm 0.5$, and $2.1 \pm 0.7$ eV{\AA} 
 for $n=1$, $2$, and $3$). 
This indicates that the Rashba-type spin-splitting of BiTeX predominantly originates from the bulk crystal structure, not the potential gradient arising from the band-bending. 
Regarding the X-dependence of Rashba parameters, the experimentally obtained values of $\alpha_{\rm{R}}$ for BiTeI ($3.9$ - $4.3$ eV\AA ) are significantly larger when compared to BiTeBr ($2.0$ - $2.1$ eV\AA ) and BiTeCl ($1.7$ - $2.2$ eV\AA ). 
Similar values and tendencies are also indicated in the series of $\alpha_{\rm{R}}^c$ estimated from the calculated subbands, as shown in Table I.
The huge spin-splitting of the CB in bulk BiTeI has been successfully explained by means of $k \cdot p$ perturbation theory \cite{Bahramy11}. 
According to this model, the SOI-induced band splitting near the $E_{\rm{F}}$, $i.e.$ CBM and valence band maximum (VBM), is strongly affected by the unusual electronic structures around its band gap.  
In BiTeX, the VBM (CBM) is commonly composed of Te $5p$ (Bi $6p$) orbitals.
The band-gap size for BiTeX, on the other hand, is known to depend on X anions: The minimum gaps realized in BiTeX are found to be $0.36$, $0.6$, and $0.7$ eV for X = I, Br, and Cl, respectively, from optical studies \cite{Leeun}. 
The narrow energy separation between CBM and VBM, combined with the large atomic SOI of Bi, enhances the spin splitting {\it via} the 
second order perturbative $k\cdot p$ Hamiltonian \cite{Bahramy11}. 
Considering these aspects, the X-dependent spin-splitting in BiTeX can be primarily understood to follow the inverse of the 
band-gap size, as observed. 

Finally, 
we investigate the effect of the 2D electron accumulation in a wider energy range, including the valence bands. 
The second-derivative ARPES images for BiTeI taken along $k_x$ direction are shown in Figs. 4(a)-(c) for $h\nu = 64$, $72$, and $80$ eV, which correspond to $k_z = \pi /c$, $1.51 \pi /c$, and $0$ $(2\pi /c)$. 
The conduction subbands with $E_{\rm{CBM}} = 0.3$ eV (red) and $0.2$ eV (blue) for $n = 1$ and $2$, similar to the laser-ARPES result in Fig. 3(a), are clearly observed. 
They are commonly observed for $h\nu = 60$ - $84$ eV, thus indicating their 2D characters. 
As for the valence bands, the peak positions of the ARPES images which are $k_z$-dependent ($k_z$-independent) are carefully separated and plotted by green (red and blue) markers.
The green markers correspond to the bulk band dispersions, which agree well with those previously obtained by the bulk-sensitive soft x-ray $h\nu$-dependent ARPES study \cite{Sakano12}.
Because of the surface sensitivity, however, the presently observed ARPES image tends to be broadened due to the $k_z$ summation. 
On the other hand, the $k_z$-independent 2D band dispersions can be clearly observed, whose shapes are strikingly different from any $k_z$ components of the bulk band dispersions. 
To account for this, we calculated the near-surface electronic structure of BiTeI in a wider energy-range. 
Figure 4(d) shows the electronic structure formed in the topmost Te-Bi-I trilayer (TL), whereas Fig. 4(e) indicates the projection of all TLs 
into the bulk. 
Surprisingly, these calculated data show the ladders of spin-split quantized subbands ($n=1$ and $2$) that 
are formed near the surface (1TL), extending in a wide energy region of the valence bands. 
By comparing with this calculation, the red (blue) markers in the ARPES image can be successfully assigned to the $n = 1$ ($n = 2$) 
subbands of the valence bands. 
The observed five sets of VB subband structures (the deepest one could not be clearly distinguished experimentally) agree well with the PS calculation.
It thus 
indicates that not only conduction-bands but also valence-bands, that are deeply bound to $\sim 4$ eV below $E_{\rm F}$,  
are well separated from the bulk and quantized near the surface as a consequence of 2D electronic confinement.


The present finding of 2D strong Rashba-field electron system, standing separately from bulk in a wide-energy range, 
thus reveals the potential of polar BiTeX for designing various interfaces with strong SOI. 
It is also worth noting that the Rashba parameter and the bulk chemical-potential are tunable in this system 
by chemical substitution and doping. 
The origin and mechanism of the near-surface band-bending for BiTeX, nevertheless, are yet unclear. 
Very recently, several studies report that accumulation (depletion) tends to  
occur at the Te- (X-) terminated surface \cite{Eremeev12, Crepaldi12, Landolt12}.
It will be interesting to elucidate the near-surface charge accumulation in polar semiconductors  
from the viewpoint of the bulk (carrier density, band structure, polarity, {\it etc.}) 
and surface/interface properties, 
for further engineering of 2D strong SOI electron systems.  \\

In summary, we investigated the electronic structure near the surface of polar semiconductors BiTeX. 
The quantized subbands with huge Rashba-type splitting were commonly observed in BiTeX, indicating  
the 2D electron accumulation layer with strong SOI robustly realized owing to the peculiar bulk structure.
The observed X-dependence of the Rashba parameter shows the importance of the band-gap size in designing the SOI field near $E_{\rm F}$. 
Together with the PS calculation, the 2D quantization was found to occur in a wide range of energy, including the valence bands deeply bound at several eV below $E_{\rm F}$.

This work was partly supported by JST, PRESTO, Grant-In-Aid for Japan, and by Funding Program for World-Leading Innovation R \& D on Science and Technology (FIRST).
H.Y.H. acknowledges support by the Department of Energy, Office of Basic Energy Sciences, Materials Sciences and Engineering Division, under contract DE-AC02-76SF00515.


\newpage

\begin{figure}[htbp!]
\caption{
(Color online) Crystal structures of BiTeI, BiTeBr (a) and BiTeCl (b). 
Brillouin zones of BiTeI, BiTeBr (c) and BiTeCl (d). 
(e) Temperature-dependent electrical resistivity of BiTeX. (f)-(h) Band dispersions of BiTeX 
obtained from first-principles calculations.
\label{fig1}
}
\end{figure}

\begin{figure}[htbp!]
\caption{
(Color online) (a) Laser-ARPES spectral image of BiTeBr along the $k_y$ direction.  (b) Constant energy cross-sectional views of the subband structure at intervals of $80$ meV. (c),(d) Depth ($z$)-dependent potential energy $V(z)$ and electron density $n(z)$ of near-surface BiTeX 
calculated by the PS method.
\label{fig2}}
\end{figure}

\begin{figure}[htbp!]
\caption{
(Color online) (a)-(c) Laser-ARPES spectral images of BiTeX. The peak positions of ARPES intensity  corresponding to $n=1$, $2$, and $3$ subbands are plotted by 
red, blue, and green markers. (d)-(f) Corresponding subband structures obtained by PS method. 
(g)-(i) Intensity mappings at $E_{\rm F}$ obtained by laser-ARPES, which correspond to the 2D Fermi surfaces near the surface. 
(j)-(l) Corresponding Fermi surfaces calculated by PS method. 
$z$-dependent potential energy and electron density in obtained are shown in Figs. 2(c) and (d).
\label{fig3}
}
\end{figure}

\begin{figure}[htbp!]
\caption{
(Color online) (a)-(c) Second-derivative ARPES images of BiTeI along the $k_x$ direction at $k_z = \pi /c$, $1.51\pi /c$, and $0$ ($2\pi /c$), respectively. The peak positions of ARPES intensity corresponding to the $k_z$-dependent bulk bands ($k_z$-independent subbands for $n=1$ and $2$) are plotted by green (red and blue) markers. 
(d),(e) Electronic structures for the topmost Te-Bi-I trilayer and 
the projection of the electronic structures from all trilayers into the bulk, obtained by PS calculation.
\label{fig4}
}
\end{figure}

\newpage

\begin{table}[htbp]
\caption{Rashba energy $E_{\rm R}$ (meV), momentum offset $k_{\rm{0}}$ ($10^{-3}$\AA $^{-1}$), experimental and
calculated Rashba parameters $\alpha_{\rm{R}}$ and $\alpha_{\rm{R}}^c$ (eV\AA )  
of the $n$-th subband dispersions in BiTeX. Numbers in parentheses represent the respective range of error.}
\begin{center}
\begin{tabular}{ccccccccc}
\hline \hline
            & \multicolumn{ 2}{c}{BiTeI} & \multicolumn{ 3}{c}{BiTeBr} & \multicolumn{ 3}{c}{BiTeCl} \\ \hline
$n$                ~&~ {$1$}   & {$2$}  ~&~  {$1$} & {$2$} & {$3$} ~&~  {$1$} & {$2$} & {$3$}   \\ \hline
$E_{\rm{R}}$       ~&~ 108(13) & 92(13) ~&~ 42(10) & 37(7) & 35(8) ~&~ 25(10) & 17(6) & 20(6)   \\ 
$k_{\rm{0}}$       ~&~ 50(10)  & 47(7)  ~&~ 43(10) & 38(6) & 34(8) ~&~  26(8) & 20(6) & 18(8)   \\ 
$\alpha_{\rm{R}}$  ~&~4.3(9)   & 3.9(8) ~&~ 2.0(7) & 2.0(5)& 2.1(7)~&~ 1.9(10)& 1.7(8)& 2.2(12) \\ 
$\alpha_{\rm{R}}^c$~&~ 3.06    & 3.02   ~&~ 1.78   & 1.67  & 1.84  ~&~ 1.28   & 1.29  & 1.78    \\ \hline \hline
\end{tabular}
\end{center}
\label{}
\end{table}

\newpage
Fig. 1 M. Sakano {\it{et al}}.
\begin{center}
 \includegraphics[height=15cm,width=15cm,keepaspectratio,clip]{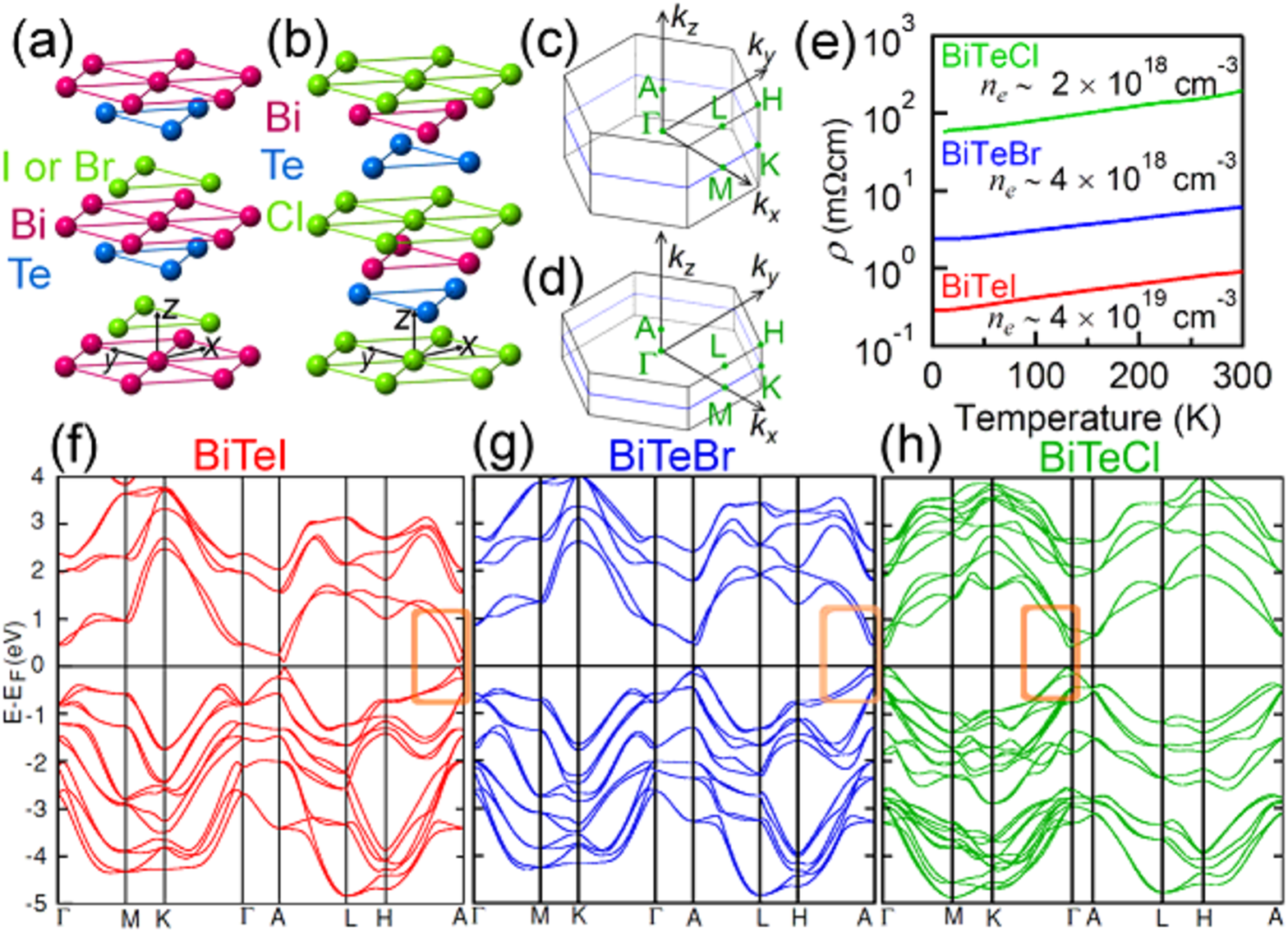}
\end{center} 

\newpage
Fig. 2 M. Sakano {\it{et al}}.
\begin{center}
 \includegraphics[height=15cm,width=15cm,keepaspectratio,clip]{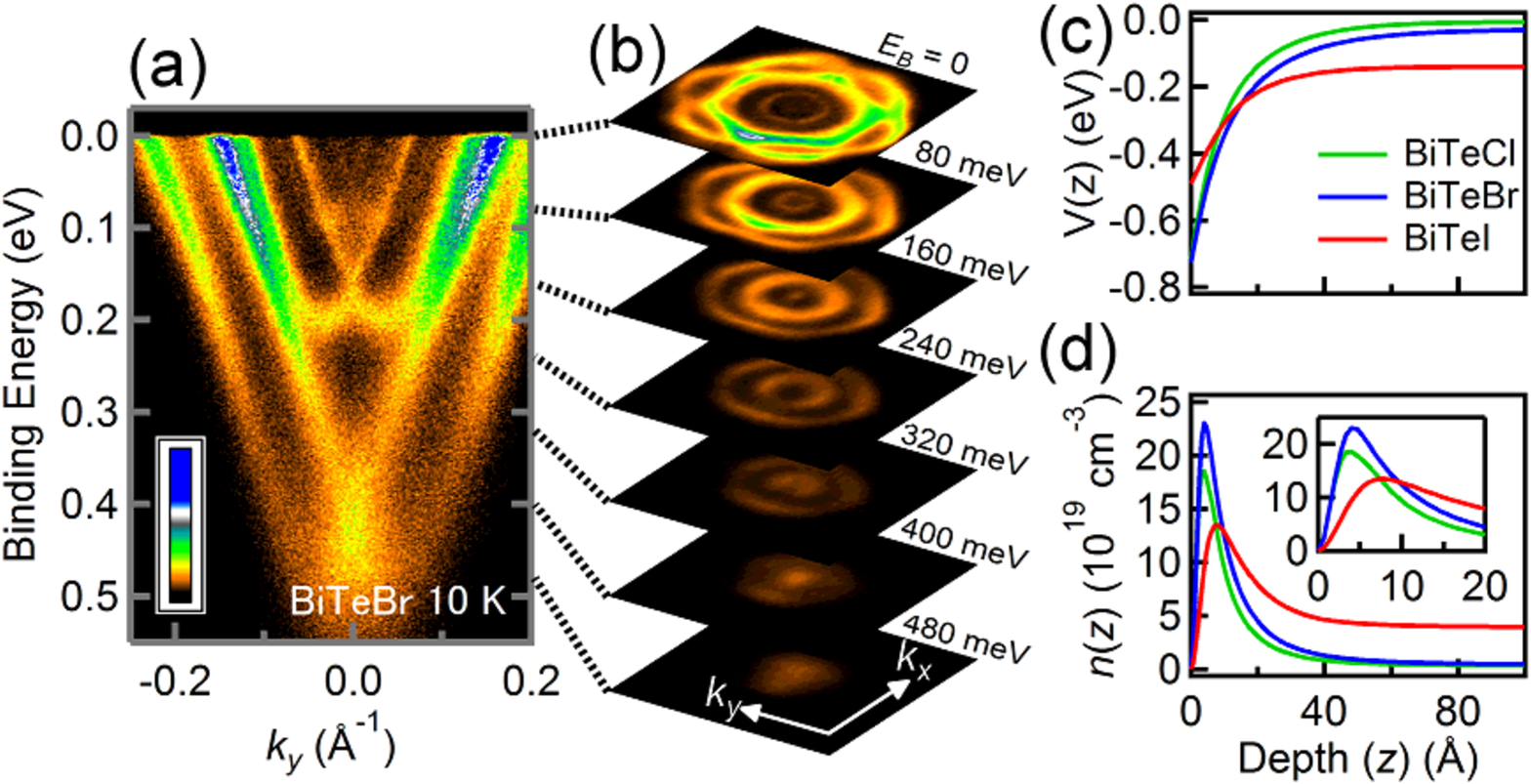}
\end{center} 

\newpage
Fig. 3 M. Sakano {\it{et al}}.
\begin{center}
 \includegraphics[height=15cm,width=15cm,keepaspectratio,clip]{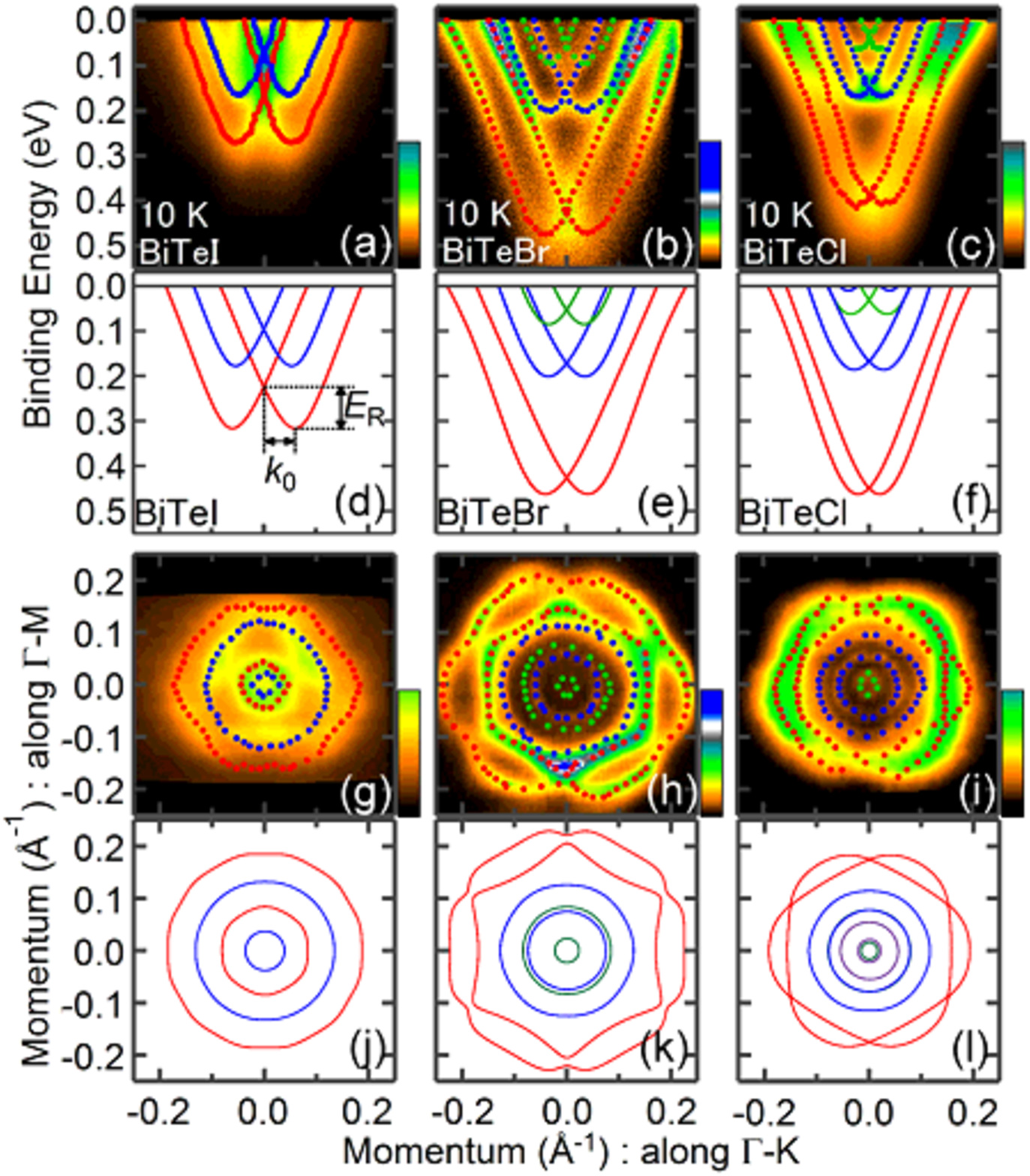}
\end{center} 

\newpage
Fig. 4 M. Sakano {\it{et al}}.
\begin{center}
 \includegraphics[height=15cm,width=15cm,keepaspectratio,clip]{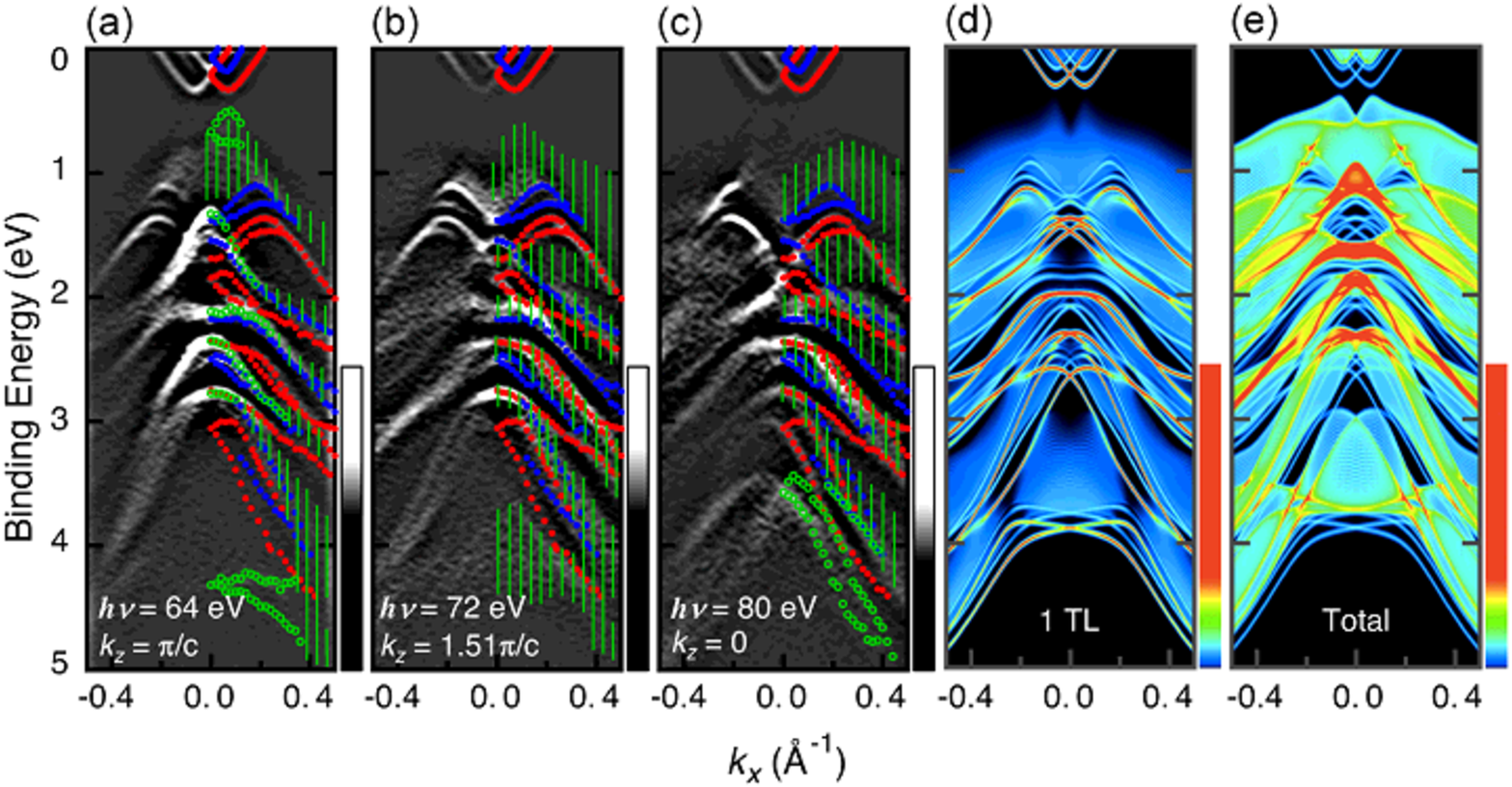}
\end{center} 

\end{document}